# Curating Safe Spaces for Emotion Disclosure during Illness Recovery on Video-Sharing Platforms

Jiaying "Lizzy" Liu[1], Yunlong Wang[2], Allen Jue[3], Yao Lyu[4], Yiheng Su[5], Shuo Niu[6], Nathan TeBlunthuis[5], Yan Zhang[5]

[1] Wee Kim Wee School of Communication and Information, Nanyang Technological University, Singapore

[2] Institute of High Performance Computing (IHPC), A*STAR, Singapore

[3] Department of Computer Science, The University of Texas at Austin, USA

[4] School of Information, University of Michigan, USA

[5] School of Information, University of Texas at Austin, USA

[6] Department of Computer Science, Clark University, USA

## 1 Introduction

The advent of video-sharing platforms (Liu & Zhang, 2024) has transformed self-expression, transcending entertainment to encompass sensitive disclosures and emotional revelations. In this submission, we discuss our study on emotional disclosure by individuals with Schizophrenia through YouTube vlogs and its implications for designing safe online spaces.

## 2 Extended Abstract

### How Schizophrenia Vloggers Construct Visual Illness Narratives on YouTube?

Understanding illness narratives in online spaces, especially for socially stigmatized diseases, has been a central topic in Human-Computer Interaction (HCI) research. Scholars have found that technology-mediated disclosures, expressions, and discourses can contribute to both the empowerment and, unfortunately, sometimes the oppression and marginalization of certain illness experiences (Feuston & Piper, 2019) and communities (Pendse et al., 2023). Thus, a primary goal for researchers and designers has been to build safe and secure online spaces where people can share illness experiences to achieve therapeutic effects, improve psychological well-being (Pennebaker, 1997), and thrive in supportive communities.

Emotion disclosure is a paramount aspect of illness narratives, given the sensitive nature of many health conditions. Studies have shown that expressing emotions about one's illness can lead to improved health outcomes and better coping strategies (Ernala et al., 2017). However, previous studies have mostly focused on text-based illness narratives on social media platforms like Twitter (Lachmar et al., 2017), Weibo (Wang et al., 2019), and Reddit (Chen & Xu, 2021). While researchers have noticed an increase in sharing illness experiences on video-sharing platforms, there is a lack of understanding of how people construct their illness narratives through videos. Videos offer unique affordances for self-disclosure, enabling various styles such as "card stories" (Misoch, 2014) or as part of gaming live streams (Mickles & Weare, 2020). These visual narratives can provide richer, more nuanced representations of lived experiences with illness than text or image alone.

An increasing number of studies have begun to delve into visual narration of mental health experiences through images and videos on platforms like Instagram (Andalibi, 2017) and YouTube (Woloshyn & Savage, 2020). The visual narration techniques employed in these videos offer a rich palette for emotional expression, far surpassing the capabilities of text or static images. Consider the "card story" phenomenon observed by Misoch (2014), where individuals bare their mental health struggles through written cards held before the camera. This seemingly simple act allows narrators to maintain a protective emotional distance while simultaneously engaging in profound self-disclosure. The visual medium also lends itself to raw, unfiltered emotional displays, as evidenced by Berryman and Kavka's (2018) examination of videos featuring individuals openly weeping or expressing anxiety on camera. Such visceral demonstrations of vulnerability challenge societal norms and push the boundaries of what's considered acceptable in public discourse about mental health.

However, individuals' practices of visual illness narration are still understudied. To address this gap in the literature, our study investigates how individuals with schizophrenia construct illness narratives through YouTube vlogs. Schizophrenia is a type of severe mental illness (SMI) that profoundly interferes with an individual's optimal functioning in major life activities. By focusing on schizophrenia, we aim to shed light on the unique challenges and experiences of individuals living with this condition. We collected 554 vlog videos posted by individuals with schizophrenia in 2022 and 2023 and examined the self-disclosure features in these videos, as well as how they influence engagement and reactions from the audience. Our research questions are as follows:

- **RQ1:** What emotions do schizophrenia vloggers disclose, and what visual narration techniques do they use?
- **RQ2:** How do the disclosed emotions and visual narration techniques influence video viewership?
- **RQ3:** How do the disclosed emotions and visual narration techniques influence support and disclosure in comments?

We reported the versatile emotion disclosure needs of Schizophrenia vloggers and their practices in creating visual narratives on YouTube. In comparison to previous studies that primarily focus on distress disclosure, we found that the vloggers expressed their fear, joy, and sadness combating the severe mental illness symptoms, recovering journeys, and disappointment of rare audience engagement and support. The vloggers use different framing and aesthetics in visual narration. We found that people express their emotions through anonymous, talk-to-camera, and in-the-moment framing. The aesthetic features of the visual disclosure vary among vlogs. We observed correlations between the valence of emotions and the visual features, specifically, in videos with more positive emotions, the visual framing tends to have better aesthetics and use in-the-moment framing. We observed 'positivity bias' and 'aesthetic bias' in audience engagement with videos, specifically, videos with in-the-moment framing and positive valences received significantly more views and supportive comments. We discussed the role of video-sharing platforms in providing safe spaces for sensitive emotional expression for both content creators and audiences and specific design implications for algorithm, and platform community-building support.

## 3 Envisioning a Safe Socio-Technical Ecosystem for Online Help-Seeking

### *3.1 Increasing the Visibility of Videos Calling for Help*

Previous studies suggest that individuals with socially stigmatized diseases or identities use online platforms as anonymous spaces where they can express negative experiences and emotions without the burden of societal expectations to present a fully functional self. These online spaces are designed to foster interactions with audiences and connect with others who share similar experiences, thus validating their personal narratives. However, our research reveals potential biases in YouTube videos, challenging the assumption that these platforms uniformly support individuals in distress. While some studies indicate that disclosing distress on social media can alleviate burdens (Zhang et al., 2021), our findings highlight a "positivity bias," where videos with higher positive valence attract significantly more views and supportive comments. This bias presents a critical challenge to the effectiveness of these platforms in supporting individuals with complex emotional needs.

### *3.2 Connection-Centered Cross-Platform Recommendations of Resources*

The diverse affordances of platforms regarding self-disclosure modalities, privacy control, and community support necessitate a more nuanced approach to resource recommendation. We propose developing an advanced, connection-centered search engine for stories and experiences. This innovative tool would prioritize peer connections over keyword-based content matching, addressing the fundamental desire of individuals with mental health concerns to connect with others and validate their experiences.

Moreover, we advocate for a paradigm shift in platform recommendation systems. Instead of defaulting to national hotlines and professional institutions, these systems should intelligently suggest informal support channels. This user-centric approach would create a more comprehensive support ecosystem, recognizing the diverse needs and preferences of individuals seeking help and connection in digital spaces.

By implementing these strategic enhancements, we can forge a safer, more inclusive digital environment that not only supports emotional disclosure but also cultivates meaningful connections for individuals navigating the complex landscape of mental health challenges. This approach promises to transform online platforms into truly supportive spaces, where even the most vulnerable voices can be heard and validated.

## References

Andalibi, N. (2017). Self-disclosure and response behaviors in socially stigmatized contexts on social media: The case of miscarriage. In *Proceedings of the 2017 CHI Conference Extended Abstracts on Human Factors in Computing Systems* (pp. 248–253). Association for Computing Machinery. https://doi.org/10.1145/3027063.3027137

Berryman, R., & Kavka, M. (2018). Crying on YouTube: Vlogs, self-exposure and the productivity of negative affect. *Convergence, 24*(1), 85–98. https://doi.org/10.1177/1354856517736981

Chen, Y., & Xu, Y. (2021). Exploring the effect of social support and empathy on user engagement in online mental health communities. *International Journal of Environmental Research and Public Health, 18*(13), Article 6855. https://doi.org/10.3390/ijerph18136855

Ernala, S. K., Rizvi, A. F., Birnbaum, M. L., Kane, J. M., & De Choudhury, M. (2017). Linguistic markers indicating therapeutic outcomes of social media disclosures of schizophrenia. *Proceedings of the ACM on Human-Computer Interaction, 1*(CSCW), 1–27. https://doi.org/10.1145/3134678

Feuston, J. L., & Piper, A. M. (2019). Everyday experiences: Small stories and mental illness on Instagram. In *Proceedings of the 2019 CHI Conference on Human Factors in Computing Systems* (pp. 1–14). Association for Computing Machinery. https://doi.org/10.1145/3290605.3300495

Lachmar, E. M., Wittenborn, A. K., Bogen, K. W., & McCauley, H. L. (2017). #MyDepressionLooksLike: Examining public discourse about depression on Twitter. *JMIR Mental Health, 4*(4), Article e43. https://doi.org/10.2196/mental.8141

Liu, J., & Zhang, Y. (2024). Modeling health video consumption behaviors on social media: Activities, challenges, and characteristics. *Proceedings of the ACM on Human-Computer Interaction, 8*(CSCW1), Article 208. https://doi.org/10.1145/3653699

Mickles, M. S., & Weare, A. M. (2020). Trying to save the game(r): Understanding the self-disclosure of YouTube subscribers surrounding mental health in video-game vlog comments. *Southern Communication Journal, 85*(4), 231–243. https://doi.org/10.1080/1041794X.2020.1798494

Misoch, S. (2014). Card stories on YouTube: A new frame for online self-disclosure. *Media and Communication, 2*(1), 2–12. https://doi.org/10.17645/mac.v2i1.16

Pendse, S. R., Kumar, N., & De Choudhury, M. (2023). Marginalization and the construction of mental illness narratives online: Foregrounding institutions in technology-mediated care. *Proceedings of the ACM on Human-Computer Interaction, 7*(CSCW2), Article 346. https://doi.org/10.1145/3610195

Pennebaker, J. W. (1997). Writing about emotional experiences as a therapeutic process. *Psychological Science, 8*(3), 162–166. https://doi.org/10.1111/j.1467-9280.1997.tb00403.x

Wang, Z., Wang, H., Zhang, Y., & Chen, R. (2019). Can people's depression level affect how they respond to related information?: Information relevance as a mediator. In *2019 4th International Conference on Communication and Information Systems (ICCIS)* (pp. 158–163). IEEE. https://doi.org/10.1109/ICCIS49662.2019.00035

Woloshyn, V., & Savage, M. J. (2020). Features of YouTube™ videos produced by individuals who self-identify with borderline personality disorder. *Digital Health, 6*, 2055207620932336. https://doi.org/10.1177/2055207620932336

Zhang, R., Bazarova, N. N., & Reddy, M. (2021). Distress disclosure across social media platforms during the COVID-19 pandemic: Untangling the effects of platforms, affordances, and audiences. In *Proceedings of the 2021 CHI Conference on Human Factors in Computing Systems* (pp. 1–15). Association for Computing Machinery. https://doi.org/10.1145/3411764.3445134